\documentstyle[prl,aps,epsf,multicol]{revtex}
\newcommand{\note }[1]{#1}
\newcommand{\eq}[1]{eq.(\ref{#1})}
\newcommand{\Eq}[1]{Eq.(\ref{#1})}

\begin{document}
\draft

\title{Temperature dependence of optical spectral weights in
quarter-filled ladder systems}

\author{ Markus Aichhorn$^{1,2}$, Peter Horsch$^1$,  Wolfgang von der Linden$^2$
and Mario Cuoco$^3$}

\address{$^1$Max-Planck-Institut f\"ur
    Festk\"orperforschung,Heisenbergstr. 1,D-70569 Stuttgart, Germany}

\address{$^2$ Institut f\"ur Theoretische Physik,
Technische Universit\"at Graz,
Petersgasse 16, A-8010 Graz, Austria}

\address{$^3$I.N.F.M. di Salerno,
Dip. Scienze Fisiche ''E.R. Caianiello'',
I-84081 Baronissi Salerno, Italy}

\date{\today}
\maketitle

\begin{abstract}
The temperature dependence of the integrated optical conductivity
$I(T)$ reflects the changes of the kinetic energy as spin and charge correlations
develop. It provides a unique way to explore experi\-mentally the
kinetic properties of strongly correlated systems.
We calculated $I(T)$ in the frame of a  $t$-$J$-$V$ model
for ladder systems,  like
NaV$_{2}$O$_{5}$, and show that the measured strong
$T$ dependence of $I(T)$ for
NaV$_{2}$O$_{5}$ can be
explained by the destruction of
short range antiferromagnetic
correlations. Thus $I(T)$ provides detailed information
about super-exchange  and magnetic energy scales.
\end{abstract}


\begin{multicols}{2}

For discrete lattice models, which are usually employed to describe
strong correlation physics in transition metal oxides,
the integrated optical conductivity (IOC) is directly proportional to the kinetic energy
\cite{baeriswyl}:
\begin{equation}\label{eq:I_T}
I_{\alpha}(T) = \int_0^{\infty} d\omega
\sigma_{\alpha}(\omega)\propto \langle-H_{\text{kin},\alpha}\rangle,
\end{equation}
which applies for finite temperatures. \Eq{eq:I_T} is even more powerful, since it holds
for different polarizations $\alpha$ individually.
This option
is particularly useful for anisotropic systems, like the quarter-filled ladder compounds
NaV$_{2}$O$_{5}$ and  LiV$_{2}$O$_{5}$
which show highly anisotropic optical spectra\cite{damascelli,golubchik,konstantinovic}.
Here the T-dependence of IOC   provides insight
into the interplay between kinetic energy and interactions.

Our work is motivated by a recent study of the optical properties of NaV$_{2}$O$_{5}$
by Presura {\it et al.}\cite{presura}, who observed a reduction of $I(T)$ by 12-14\%
between $4$K and room temperature.
They proposed a fitting formula for the optical conductivity (integrated up to 2.25 eV)
$  I(T)/I(0)=(1-f \exp (-E_0/T))$
with $f\sim 0.35 (0.47)$ and $E_0\sim 286 (370)$K
for $a (b)$ polarization, respectively,
predicting a reduction of almost 50\% for $b$-polarization at
several hundred degrees centigrade.
As the main absorption is near 1 eV one may wonder about the origin of such a dramatic
change of kinetic energy at low temperatures.

A key feature of NaV$_{2}$O$_{5}$,
closely related to its charge dynamics\cite{atzkern}, is the 3D
charge ordering transition at 34K\cite{ohama}
which is accompanied by the opening of a spin
gap\cite{isobe}. The precise structure of the low-T phase is still under
dispute\cite{luedecke}.
As  discussed below, the charge fluctuations of a single ladder
can be mapped onto the Ising model in a transverse field (IMTF)\cite{khomskii}.
It has been pointed out in Refs.\cite{khomskii,cuoco} that the parameters
for  NaV$_{2}$O$_{5}$ are such that the IMTF is close to its quantum critical
point. The corresponding soft charge excitations appear at $q_b=\pi$ and
therefore do not directly contribute to $\sigma(\omega)$. Yet it was shown
in Ref.\cite{cuoco} that when including spin degrees of freedom
the soft charge excitations
contribute a small absorption continuum in $\sigma_a(\omega)$
within the charge gap, and may explain the anomalous absorption observed
by Damascelli {\it et al.}\cite{damascelli}.
Presura {\it et al.}\cite{presura} suggested that
the low energy excitations at  $q_b=\pi$ may
also explain the 30 meV activation energy for charge transport\cite{hemberger},
which is surprisingly low
in view of the 0.8 eV optical gap. Furthermore it was conjectured\cite{presura}
that these excitations may cause the T-dependence of IOC's
and hence $E_0$ in Presura's fit should measure the charge gap.

The picture which evolves from our calculations is different.
The kinetic energy of the IMTF does not show any significant
temperature dependence which could be attributed to $E_0$. The temperature scale of
the variation of the kinetic energy  is set by the bare Coulomb interaction.
The $t$-$J$-$V$ model, however, in agreement with the experimental data, displays a dramatic decrease of kinetic energy
 in the range
$0.2 J< T < J$, where $J$ is the exchange integral\cite{horsch,yushankhai} (order 0.1eV)
of the effective 1D Heisenberg model which describes the spin dynamics of
NaV$_{2}$O$_{5}$\cite{isobe}.
We attribute the large decrease of $I(T)$ to the destruction
of short range AF correlations by thermal population of local triplet excitations.
Thereby IOC's allow to measure the super-exchange contribution to $E_{\text{kin}}$.

At quarter filling, NaV$_{2}$O$_{5}$ can be described by a $t$-$J$-$V$
model\cite{smolinski,ohta,riera}, as the hopping
matrix elements $t_{ij}$ between vanadium $d_{xy}$ orbitals are
small compared with the intra-atomic interaction ($U \sim 4$ eV):
\begin{eqnarray}\label{eq:tVJ_model}
H &=& -\sum_{\langle i,j\rangle,\sigma} t_{ij} (\tilde{c}^{\dagger}_{i,\sigma}
\tilde{c}_{j,\sigma}+H.c.)\nonumber\\
&+&\sum_{\langle i,j\rangle} J_{ij} ({\bf S} _{i} \cdot {\bf S}_{j}
-\frac{1}{4} n_{i} n_{j})
+\sum_{\langle i,j\rangle} V_{ij}~n_{i} n_{j}.
\label{tJV}
\end{eqnarray}
Elimination of local double occupancies\cite{note0}
yields the Heisenberg term with $J_{ij}=4 t^2_{ij}/U$ and
constrained electron creation operators
$\tilde{c}^{\dagger}_{i,\sigma}={c}^{\dagger}_{i,\sigma}
(1-n_{i,-\sigma})$, which enter the densities
$n_{i}=\sum_{\sigma}\tilde{c}^{\dagger}_{i,\sigma}\tilde{c}_{i,\sigma}$
in the Coulomb repulsion ($V_{ij}$) between neighbors (Fig.1).
${\bf S} _{i}$ is the spin-$\frac{1}{2}$ operator at site $i$. The sums
are over spin $\sigma=\uparrow,\downarrow$ and all neighbor bonds $\langle i,j\rangle$ on the
trellis lattice, depicted in fig.1. Typical parameters values are included in Fig.3.
%
\begin{figure}
\epsfxsize=45mm
      \centerline{\epsffile{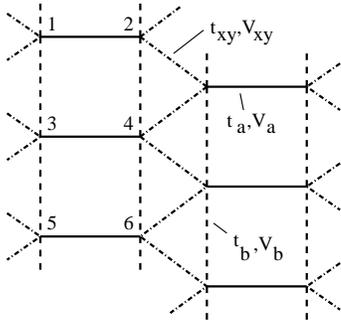}}
\caption {\label{fig_a}
Trellis lattice describing the positions of vanadium ions in the (a,b)-plane of
$\alpha$'-NaV$_2$O$_5$.
}
\end{figure}
We begin with the investigation of charge fluctuations in the spinless case.
Due to the interactions $V_{ij}$ this is a nontrivial problem, which can be simplified
to a single ladder model without changing the essence of the problem,
provided $V_{xy}$ and $t_{xy}$ are not too large.
For large $U/|t_{ij}|$ and $V_a/|t_{ij}|$
the relevant subspace of one electron per rung can be represented by pseudospin
operators ${\bf T}_r$, where the eigenvalues $\pm \frac{1}{2}$ of
$T_r^{z}$ correspond to the left/right position of the electron
within the rung $r$, and one obtains the one-dimensional (1D) IMTF\cite{khomskii}:
\begin{eqnarray}\label{eq:ising}
H_{\text{ladder}}=-2 t_{a}\sum_{r} T^{x}_{r} +2 V_{b} \sum_{r} T^{z}_r
T^{z}_{r+1}. \label{ITF}
\end{eqnarray}
Here the Ising interaction, due to Coulomb repulsion, favors zigzag charge correlations on
the ladder, while the kinetic energy $E_{\text{kin}}$ ($\propto t_a$)
appears as a transverse field opposing these correlations.

The 1D IMTF is exactly solvable\cite{lieb}.
The kinetic energy,  expressed in terms of the dimensionless
parameter $h=2 t_a/V_b$, is given by
\begin{equation}\label{eq:E_kin_ex}
\frac{E_{\text{kin}}}{V_b}=\frac{h}{2\pi}\int^{\pi}_0 dk \frac{\cos{k} +h}{\epsilon (k)}
\tanh (\frac{\beta V_b \epsilon (k)}{2}),
\end{equation}
where $\beta=1/k_{B}T$ and
$\epsilon (k)=\sqrt{h^2 + 2h\cos{k} +1}$ is the quasiparticle dispersion
which becomes soft at $k=\pi$ as $h\rightarrow 1$
(i.e. $E_0=V_b \epsilon(\pi)=|2 t_a-V_b|$).

Fig.2 shows the temperature dependence of the exact results
for $E_{\text{kin}}$ and the pseudo-spin
correlation functions (CF)  $P_r =\frac{1}{N}\sum_{r'} \langle
T^z_{r'} T^z_{r'+r}\rangle$ for nearest and next-nearest neighbors ($r=1,2$).
Obviously,
the temperature dependence of $E_{\text{kin}}$ and  the CF's is dictated by the
bare interaction $V_b$ as only T-scale, provided $t_a \gg\!\!\!\!\!\!\!\big/\quad V_b$ . The CF's decrease gradually with increasing
$h=2 t_a/V_b$, and vanish only in the limit $h\rightarrow \infty$.
We note one peculiarity of the exact solution: for $h<1$ the magnetization
$\langle T_r^x\rangle \propto E_{\text{kin}}$ is not strictly monotonic in temperature.
Here $\langle T_r^x\rangle$ profits from a decrease of  $z$-correlations.
In the high temperature limit one finds $E_{\text{kin}}\sim h/T$.


From Fig.2 we conclude that the charge-only model cannot explain the decline of $I(T)$
below room temperature.
To study the problem including spin there are two alternatives: (i) the complete
spin-pseudospin model\cite{khomskii,cuoco} and (ii) the $t$-$J$-$V$ model
which we shall choose.
\vspace*{3mm}
\begin{figure}
\epsfxsize=.7\columnwidth
      \centerline{\epsffile{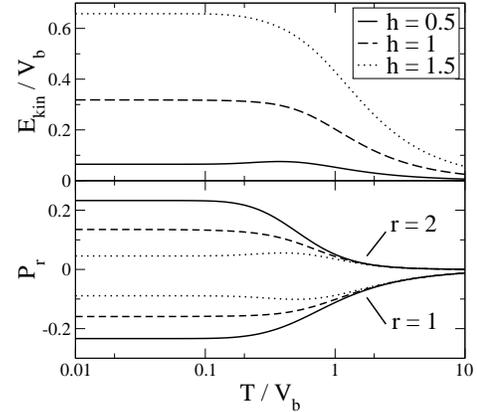}}
\vspace*{2mm}
\caption {\label{fig_b}
Temperature dependence of the kinetic energy (top)  and pseudospin CF
derived from the exact solution of the
IMTF for $h=2t_a/V_b=0.5, 1.0$, and $1.5$.
}
\end{figure}
For the calculation of the optical spectra and IOC's for the $t$-$J$-$V$ model
we employ, as in Ref.\cite{cuoco}, the finite temperature Lanczos Technique
developed by Jaklic and Prelovsek\cite{prelovsek}. The exact diagonalization (ED) was
performed for a $4\times4$ site system.
As we are studying an insulator, we can restrict the calculation of
$\sigma_{\alpha}(\omega)$ to the finite
frequency response given by the Kubo formula
\begin{eqnarray}\label{eq:Kubo}
\sigma_{\alpha}(\omega)=\frac{1-e^{-\beta \omega}}{\omega}
\text{Re} \int_0^{\infty} d\tau e^{i \omega \tau} \langle j_{\alpha}(\tau)j_{\alpha}\rangle ,
\end{eqnarray}
where
$j_{\alpha}=i \sum_{\langle i,j\rangle,\sigma} t_{ij} R^{ij}_{\alpha}\;
(\tilde{c}^{\dagger}_{i,\sigma} \tilde{c}^{\phantom{\dagger}}_{j,\sigma}-hc)$ is
the $\alpha(=a,b)$  component of the current operator and
$R^{ij}_{\alpha}$ denotes the vector connecting
sites $i$ and $j$. In the following, energies and temperature are given in eV.

Figure 3 displays the temperature variation of $I(T)$ for the
$t$-$J$-$V$ model for several  sets of parameters on a log-log scale.
The data show a strong decrease at low temperature, which is
particularly pronounced for $b$-polarization, while the decrease of the kinetic energy at
high temperatures \hbox{$(T\gtrsim V_b)$} is similar to that discussed for the pseudo-spin
(charge only) model. The low-T variation is controlled by the magnetic exchange
$J \propto t^2_b/V_b$ as can be seen by comparing the data for $t_b=0.1$ and 0.2 eV.
Furthermore, the results for $t_{xy}=0$ and $t_{xy}=0.15$ are similar
(for $t_b=0.2$)\cite{note1} which is consistent
with the suppression of super-exchange related to $t_{xy}$, as discussed in Ref.\cite{horsch}.

The near-neighbor spin CF's $S_{1n}$
for $t_b=0.2$ and $t_{xy}=0.15$ are shown in Fig.4. The comparison with the
kinetic energy for these parameters (open circles in Fig.3)
clearly illustrates that the low temperature decrease in $E_{\text{kin}}$ is
due to the loss of short-range AF spin-correlations,
whose $T$-variation is controlled by the magnetic exchange $J$.
The largest spin CF's at low temperatures are $S_{14}$ and $S_{15}$
consistent with zigzag charge correlations, i.e. large  $C_{14}$ and $C_{15}$.
As discussed along with
Fig.2, the short-range charge correlations, however, exist
up to higher temperatures, determined by $V_b$.

\begin{figure}[hb]
\epsfxsize=74mm
      \centerline{\epsffile{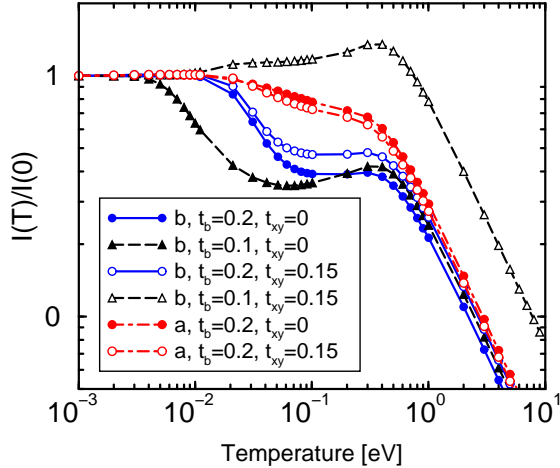}}
\caption {\label{fig_c} IOC for $a$- (dot-dashed) and $b$-polarization
(solid and dashed lines). Results are given for both $t_{xy}=0.15$
(open) and  $t_{xy}=0$ (filled symbols). Common parameters are
$t_a=0.4$, $V_a=V_b=0.8$ and $V_{xy}=0.9$ eV. The set $t_b=0.2$
and $t_{xy}=0.15$ corresponds to spectra shown in
Ref.\protect\cite{cuoco} . }
\end{figure}
\begin{figure}[hb]
\epsfxsize=65mm
      \centerline{\epsffile{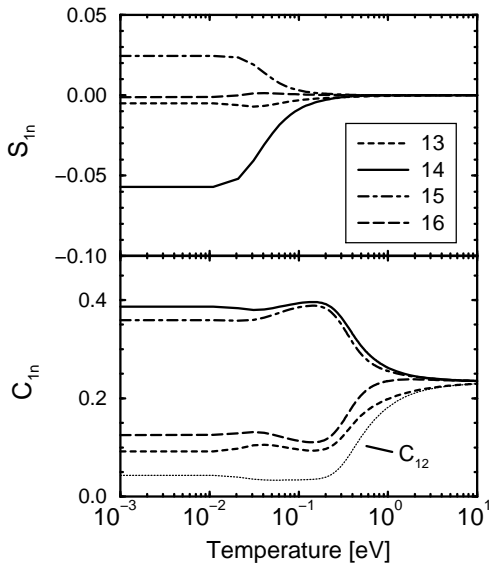}}
\caption {\label{fig_d}
Spin- $S_{1n}=\langle S^z_1S^z_n \rangle$ (top) and charge-correlations
$C_{1n}=\langle n_1 n_n \rangle$ (bottom), (see Fig.1),
as function of temperature ($t_b=0.2$, $t_{xy}=0.15$).
}
\end{figure}
Finally, the smallness of $C_{12}$ below $T\lesssim 0.4$ supports the validity of the
pseudo-spin model, where only one electron
per rung is allowed for. We also note that the data for $C_{1n}$ are consistent with
the pseudo-spin correlation $P_r$, shown in Fig. 2.

We emphasize that the kinetic energy is essentially
determined by short-range  spin-correlations.
In the insulating state $E_{kin}$ has two major contributions
which lead to the two distinct T-scales in Fig.3:
(i) valence fluctuations, which have been discussed already in the context of
the pseudo-spin model, and
(ii) virtual
excitations, which give rise to magnetic
super-exchange
whose $T$-variation is controlled by  $J$.

To  understand the low-T behavior  of IOC's it is
revealing to study a two-rung system including all terms in
\Eq{eq:tVJ_model}, except $t_{xy}$ and $V_{xy}$, and to employ the
spectral representation of $I(T)$ (\eq{eq:I_T}):
\begin{equation}\label{eq:I_T_spectral}
I(T)\!\! =\!\! \frac{\pi}{Z}\sum_{n} \;e^{-\beta
E_n}\!\!\!\!\!\sum_{m\atop E_m>E_n}\!\!\!\! \frac{1-e^{-\beta
(E_m-E_n)}}{E_m-E_n}\; |\langle m|j_{\alpha}|n\rangle|^2\!\!,
\end{equation}
where the summation is over eigenstates of $H$ and $Z$ is the
partition function. The eigenstates are characterized by
$(S,p_a,p_b)$, i.e. total spin (S=singlet/triplet) and parity
$p_{a/b}$ in a/b- direction. The level scheme is sketched in
Fig.5. The ground state $(s,+,+)$ is a fully symmetric
singlet. The lowest excited states are the
triplet states $(t,+,-)$ and $(t,-,+)$, where $(t,+,-)$ is lower
in energy, as it is composed of bonding and anti-bonding orbitals
in a-direction, while  $(t,-,+)$ consists of orbitals in
b-direction and lies already $\approx 400meV$ above the ground state. Its
thermal population in \eq{eq:I_T_spectral} can be neglected at low $T$.
The difference between the two polarizations is due to the fact that
the current operators $j_{\alpha}$ conserve the symmetries
$(S,p_a,p_b)$, except for $p_\alpha\to-p_\alpha$. 
One finds that
${j}_b$ has non vanishing matrix elements
between the ground state and the first excited singlet
with $(s,+,-)$, whereas the matrix elements involving $(t,+,-)$
vanish. This has the important consequence that the triplet state
$(t,+,-)$ contributes to  $Z$ but not to the
numerator of the Kubo formula (Eq.(6)), leading to
\begin{eqnarray}
\label{eq:fit_b}
I_b(T) &\propto & \frac{1-e^{-\beta\Delta E_s}}
        {\Delta E_s}\frac{1}{1+3 e^{-\beta E_{st}^0}}.
\end{eqnarray}
Here the factor 3 reflects the degeneracy of the triplet and
$\Delta E_s$ is the energy difference between the lowest singlet states.
The situation is different for $a$-polarization, where
also $(t,-,+)$ contributes, resulting in an extra term
\begin{eqnarray}
\label{eq:fit_a}
I_a(T) &\propto & \frac{1-e^{-\beta\Delta E_s}}{\Delta E_s (1+3 e^{-\beta E_{st}^0})} +
       3 \kappa  \frac{1-e^{-\beta\Delta E_t}}{\Delta E_t  (3+e^{\beta E_{st}^0}) },
\end{eqnarray}
where $\kappa$ is a ratio of matrix elements.
In both cases the low-T behavior is governed by local singlet-triplet excitations of energy
$E^0_{st}$ and hence by the exchange integral
$J (=E^0_{st})$ of the effective 1d Heisenberg model
$H=J \sum {\bf S}_r{\bf S}_{r+1}$\cite{horsch,smolinski}.
The influence is however different for a- and b-polarization.
The T-dependence via the optical excitation energies $\Delta E_{s}$ and $\Delta E_{t}$ ($\sim 1$ eV) is marginal
at low $T$.

We infer that also in extended systems, the low-T behavior is due to
local singlet-triplet excitations
with energy, estimated from the two-rung formula,
\hbox{
$\tilde{E}^0_{st} =
4t_b^2/[(V_a+V_b)\sqrt{1+4(t_a/ V_b)^2}]$} for $V_a,V_b\gg t_a, t_b$.
For the standard parameters and $t_b=0.2$, $\tilde{E}^0_{st}\!\!\sim\!\!70$meV.

As the results of the two-rung system are mainly
determined by symmetry arguments and energy-scale separation,
we consider the expressions in Eqs.(7,8)
as generic results with system-dependent parameters:
$E^0_{st}$, $\kappa$, $\Delta E_s$ and $\Delta E_t$.
We have determined
these parameters by fitting the ED-data of the 4x4 system (Fig.3) resulting in
$E^0_{st}\sim 76\; (61)$ meV for a (b) polarization, respectively,
and $\kappa\sim 0.54$.
The fit-value for $E^0_{st}$ agrees well with the estimate $\tilde{E}^0_{st}$.
The fit-curves along with the ED-data are shown in Fig.5.
\begin{figure}
\epsfxsize=80mm
      \centerline{\epsffile{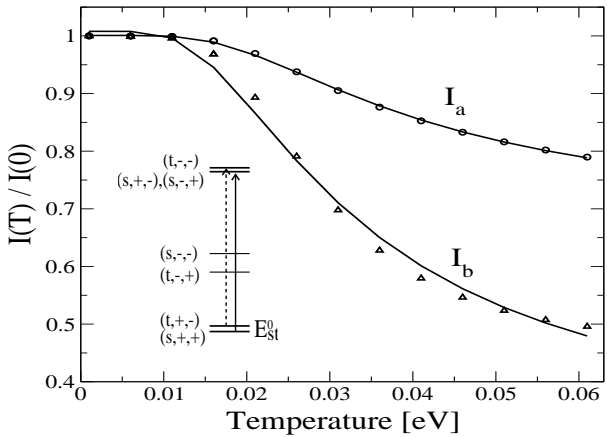}}
\vspace*{2mm}
\caption {\label{fig_e}
IOC for standard parameters with $t_b=0.2$ and
$t_{xy}=0.15$ eV: fit (solid lines) as described in the text and
ED from Fig.3 (symbols).
Inset: sketch of optical excitations.
of the two-rung model. For $b$-polarization the triplet transition
matrix element is zero (dashed).
}
\end{figure}
A similar fit of the experimental data in \cite{presura}
yields $J\sim 39$ meV, consistent with values obtained from other
experiments, e.g.  $J\sim 48.2$\cite{isobe} and 37.9 meV
\cite{weiden,yosihama,yushankhai}. According to  $\tilde{E}^0_{st}$ we estimate
$t_b\sim 0.15$, \note{like in} \cite{horsch,smolinski} instead of 0.2 eV used in this study and in
Ref.\cite{cuoco}.

Finally one may wonder about the change of $E_{kin}$ at the 34 K transition\cite{isobe}.
As this charge order transition indicates the onset of 3D long-range order, there is 
not necessarily a large change of short range correlations to be expected.
This view is consistent with the small change of $I(T)$ at 34 K in Presura's data.

In conclusion, we have shown, based on the exact CF's of the IMTF,
i.e. ignoring spin degrees of freedom,
that the low energy excitations associated with the
(charge) correlations, do not lead to a substantial T-dependence of IOC's.
However, for the quarter-filled $t$-$J$-$V$ model we find, in agreement with
experimental data, a large decrease of IOC's in the temperature range
$0.2 J<T< J$. This change in kinetic energy is magnetic in origin, and
is explained by the destruction of short-range spin-correlations as
T is increased. 
We argue therefore, that the observed T-dependence of optical
spectral weights in NaV$_{2}$O$_{5}$
is explained by the decrease of the magnetic super-exchange.
This underlines the potential of optical experiments to
infer the magnetic exchange $J$ from the T-dependence of the spectra
and to explore
the changes of the kinetic (and super-exchange) energy
in correlated systems (particularly in systems with small $J$),
which cannot be achieved otherwise.

It is a pleasure to thank M. Konstantinovi\'c, R. K. Kremer, Y. Ueda and V. Yushankhai
for useful discussions.

\end{multicols}


\begin{thebibliography}{99}
\bibitem{note0} We note that in the quarter-filled case the
magnetic properties follow from the interplay of `$t$' and `$V$' terms in $H$
and not from the `$J$' term, as $J_{ij}=4t^2_{ij}/U$ is small.
\bibitem{note1} For small $t_b=0.1$ this is no longer the case and the structure
of the ground state is essentially changed and determined by $t_{xy}$.
\end{thebibliography}
\end{document}